\RequirePackage{ifpdf}
\ifpdf 
\documentclass[pdftex]{sigma}
\else
\documentclass{sigma}
\fi

\begin{document}

\allowdisplaybreaks

\renewcommand{\PaperNumber}{043}

\renewcommand{\thefootnote}{$\star$}

\FirstPageHeading

\ShortArticleName{A Journey Between Two Curves}

\ArticleName{A Journey Between Two Curves\footnote{This paper is a contribution to the Proceedings
of the Workshop on Geometric Aspects of Integ\-rable Systems
 (July 17--19, 2006, University of Coimbra, Portugal).
The full collection is available at
\href{http://www.emis.de/journals/SIGMA/Coimbra2006.html}{http://www.emis.de/journals/SIGMA/Coimbra2006.html}}}

\Author{Sergey A. CHERKIS~$^{\dag^1\dag^2}$}

\AuthorNameForHeading{S.A. Cherkis}
\Address{$^{\dag^1}$~School of Mathematics, Trinity College Dublin,  Ireland}
\EmailDD{\href{mailto:cherkis@maths.tcd.ie}{cherkis@maths.tcd.ie}}

\Address{$^{\dag^2}$~Hamilton Mathematics Institute, TCD, Dublin, Ireland}

\ArticleDates{Received October 31, 2006, in f\/inal form February
25, 2007; Published online March 11, 2007}

\Abstract{A typical solution of an integrable system is described in terms of a holomorphic curve and a line bundle over it.   The curve provides the action variables while the time evolution is a linear f\/low on the curve's Jacobian.  Even though the system of Nahm equations is closely related to the Hitchin system, the curves appearing in these two cases have very dif\/ferent nature.  The former can be described in terms of some classical scattering problem  while the latter provides a solution to some Seiberg--Witten gauge theory.  This note identif\/ies the setup in which one can formulate the question of relating the two curves.}

\Keywords{Hitchin system; Nahm equations; monopoles; Seiberg--Witten theory}

\Classification{53C28; 53C80; 70H06; 81T30}

\section{Introduction}
The Nahm Equations  \cite{Nahm:1979yw} and the Hitchin Equations \cite{HitchinSystem1, HitchinSystem2} are two integrable systems of equations that play increasingly important role in string theory and gauge theory literature.  Solutions of Seiberg--Witten gauge theories in four  \cite{Donagi:1995cf, Kapustin:1998pb} and in three dimensions \cite{Seiberg:1996nz}, dynamics of a f\/ive brane on a holomorphic curve \cite{Klemm:1996bj, Witten:1997sc},  D-brane dynamics  \cite{Chalmers:1996xh, Hanany:1996ie}, aspects of the Generalized Geometry~\cite{Hitchin:2005in}, physics version of Langlands duality~\cite{Kapustin:2006pk}, domain walls \cite{Hanany:2005bq}, topological Yang--Mills and nonlinear Schr\"odinger~\cite{Gerasimov:2006zt}, to name a few,  can be described in their terms.

The two systems are closely related.  In fact, one can think of the Nahm equations as a limit of those of Hitchin.  As integrable systems each can be solved in terms of a holomorphic curve and a point or a linear f\/low on its Jacobian. Yet, these two solutions are tantalizingly dif\/ferent.

Each system can be mapped to a dif\/ferent problem by the Nahm transform. The Nahm transform on a four-torus is presented beautifully in \cite{DK}; a general review of the Nahm transform can be found in \cite{Survey}.  A solution to  the Nahm equations is mapped to a monopole \cite{NahmTransform, Nahm:1981nb}, i.e. solutions of the Bogomolny equation. In turn, a solution to the Hitchin equations def\/ined on a torus is mapped to a doubly-periodic instanton \cite{Jardim:1999dx}.    A reduction of the instanton equation leads to the   Bogomolny equation.  Since instanton  and Bogomolny equations are, naturally, also integrable, one can f\/ind the  respective curves describing their solutions.  Yet again, the two curves dif\/fer drastically from each other.

In the context of monopoles and the Hitchin system respectively both curves ${\bf S}$ and ${\bf \Sigma}$ were introduced by Hitchin and both are often referred to as Hitchin Spectral Curves. To distinguish these  curves we shall refer to ${\bf S}$ as the {\em Twistor Curve} and to ${\bf \Sigma}$ as the {\em Higgs Curve}.  The two are of the very dif\/ferent nature, nevertheless, they are closely related.  As we hope to demonstrate here, this relation is deeper than originating from the same mind.

\section{The physics of the Higgs Curve ${\bf \Sigma}$ and the Twistor Curve ${\bf S}$}
Our motivation for posing the problem comes from the deep signif\/icance of both curves in the study of gauge theories with eight supercharges.   A solution to this problem would also provide a glimpse at the nonperturbative dynamics of the M theory f\/ive-brane.

\subsection[${\bf \Sigma}$ in Seiberg-Witten theory]{${\bf \Sigma}$ in Seiberg--Witten theory}
The solution of the ${\cal N}=2$ gauge theory found by Seiberg and Witten \cite{Seiberg:1994rs,Seiberg:1994aj} is formulated in terms of an auxiliary curve.   This is a holomorphic curve corresponding to a choice of a~vacuum of the theory.  It is referred to as the Seiberg--Witten curve.    As demonstrated in \cite{Donagi:1995cf}, Seiberg--Witten curve is exactly the Higgs Curve ${\bf \Sigma},$ as outlined in Section \ref{section:HiggsCurve} below.

M theory  signif\/icance of the Seiberg--Witten curve was uncovered in \cite{Klemm:1996bj, Katz:1996fh, Witten:1997sc}. In \cite{Witten:1997sc} the curve~${\bf \Sigma}$ emerges as a curve on which the M theory f\/ive-brane is wrapped.  The f\/ive brane dynamics is equivalent to that of a type IIA string theory brane conf\/iguration of Chalmers--Hanany--Witten~\cite{Chalmers:1996xh} and~\cite{Hanany:1996ie}.  In turn, the low energy dynamics of this brane conf\/iguration is described by a Seiberg--Witten gauge theory.  Thus, what appeared to be an auxiliary curve of Seiberg and Witten acquired new meaning as the shape of the M theory f\/ive-brane.

Another path to this M theory interpretation was discovered earlier in \cite{Klemm:1996bj, Katz:1996fh}.  In this Geometric Engineering scenario the Seiberg--Witten gauge theory emerges from the compactif\/ication of the IIA string theory on a Calabi--Yau space with a prescribed singular structure along a~curve in it. After local T-duality this Calabi--Yau space is mapped to a f\/ive-brane wrapped on a~Seiberg--Witten curve.

\subsection{Three-dimensional Gauge theories and ${\bf S}$}

The gauge theory relevant to this section is ${\cal N}=4$ supersymmetric gauge theory in three dimensions.   The case of pure ${\cal N}=4$  Yang--Mills was considered in \cite{Seiberg:1996nz}, where the geometry of its moduli space of vacua is identif\/ied with the Atiyah--Hitchin moduli space \cite{AtiyahHitchin}.    Chalmers, Hanany, and Witten in
\cite{Chalmers:1996xh} and \cite{Hanany:1996ie} have identif\/ied the brane conf\/iguration that allows one to associate the above quantum f\/ield theory problem with a problem of the classical dynamics of monopoles.  It also makes the relation to the corresponding Nahm data transparent.  It even allows one to re-derive the Nahm transform from the string theory \cite{Diaconescu}.  Using these ideas one can identify the exact monopole problem solving three-dimensional ${\cal N}=4$ QCD
\cite{Cherkis:1997aa}.  One can also derive the K\"{a}hler potential on the moduli space
\cite{Cherkis:1998hi}.
The auxiliary curve ${\bf S}$ plays central role in  this derivation.  Moreover, using the information encoded in ${\bf S}$ (and a bundle ${\bf\cal L}\rightarrow{\bf S}$) one can obtain the explicit metric on the moduli space
\cite{Cherkis:2003wk}. From the point of view of the original quantum gauge theory this metric contains all of the nonperturbative corrections.

The physical signif\/icance of the twistor curve ${\bf S}$ is much less clear.  It is also much harder to f\/ind it explicitly, since one has to satisfy the transcendental constraint, as stated in Section \ref{section:Constraint}.  However, the curve ${\bf S}$ contains complete information about the moduli space, as well as about the solutions  of the associated monopole and Nahm problems themselves.

\subsection[${\bf \Sigma}$ in Seiberg-Witten theories on ${\mathbb R}^3\times S^1$]{${\bf \Sigma}$ in Seiberg--Witten theories on $\boldsymbol{{\mathbb R}^3\times S^1}$}

Supersymmetric ${\cal N}=2$ gauge theories on a space with one compact direction considered in  \cite{Seiberg:1996nz} are rather revealing, since they interpolate between the four- and three-dimensional theories.  From the geometric point of view, they also have much more interesting moduli spaces of vacua.

Applying the techniques of the previous section one can identify the Chalmers--Hanany--Witten type brane conf\/iguration corresponding to the quantum gauge theory.  From this one can read of\/f the corresponding Hitchin system
\cite{Cherkis:2000ft}
and via T-duality obtain the periodic monopole description
\cite{Cherkis:2000cj}.
Both lead to the curve ${\bf \Sigma}$ which is the Seiberg--Witten curve of the quantum gauge theory problem.

Most importantly, the curve ${\bf \Sigma}$ is relatively easy to f\/ind explicitly.  It contains a wealth of information about the Seiberg--Witten gauge theory on ${\mathbb R}^3\times S^1;$ however, it is far from suf\/f\/icient to know ${\bf \Sigma}$ in order to f\/ind the ef\/fective low energy theory in this case. Physically, the main dif\/ference between the quantum f\/ield theory on ${\mathbb R}^4$ and a theory on ${\mathbb R}^3\times S^1$ is that in addition to the instanton corrections the latter also has monopole corrections, since in that case a self-dual conf\/iguration independent of $S^1$ has f\/inite action.  The Seiberg--Witten curve ${\bf \Sigma}$ accounts for all perturbative and instanton corrections to the ef\/fective theory, but not the monopole contributions.

Let us comment on the M theory signif\/icance of the problem we suggest. As explained in~\cite{Witten:1997sc}, in the extreme infra-red the ef\/fective f\/ield theory of a Seiberg--Witten theory is given by the low energy dynamics of an M theory f\/ive-brane wrapped on a curve~${\bf \Sigma}$.  The reduction of the M theory action produces the Seiberg--Witten solution \cite{Howe:1997eu}.   However, if the gauge theory is considered on ${\mathbb R}^3\times S^1$ the f\/ive brane dynamics receives membrane instanton corrections. In this case the f\/ive-brane wraps ${\bf \Sigma}\times S^1$ and for some one-cycle  $\Gamma\subset{\bf \Sigma}$ an instanton membrane of f\/inite volume can have $\Gamma\times S^1$ as its boundary.  Such a membrane instanton produces corrections to the f\/ive-brane dynamics.  At low energies, these can be interpreted as monopole corrections in the corresponding gauge theory.

The main purpose of this paper is to highlight the relation between ${\bf S}$ and ${\bf \Sigma}.$
Seiberg--Witten theories on a space with one compact direction provide a perfect framework to pose this question.  In physics terms, a vacuum of such a theory def\/ines a curve ${\bf \Sigma},$ it also def\/ines what we call a~Twistor Curve~${\bf S}_p.$  Thus, we expect these two to be related.  Moreover, we expect the family of ${\bf S}_p$ curves  to contain the complete information about the moduli space of vacua.

\section{Two Descriptions of the Twistor Curve ${\bf S}$: \\ monopoles and Nahm equations}

There is a well established relation, discovered by Werner Nahm in \cite{Nahm:1979yw, NahmTransform, Nahm:1981nb}, between solutions $(A, \Phi)$ of the Bogomolny equation
\begin{gather}\label{BogomolnyEq}
\star F=D\Phi
\end{gather}
and solutions of the system of the Nahm equations
\begin{gather}
\frac{d}{ds} T_1(s)+i[T_0(s),T_1(s)]=-i[T_2(s), T_3(s)],\nonumber\\
\frac{d}{ds} T_2(s)+i[T_0(s),T_2(s)]=-i[T_3(s), T_1(s)],\label{NahmEq}\\
\frac{d}{ds} T_3(s)+i[T_0(s),T_3(s)]=-i[T_1(s), T_2(s)].\nonumber
\end{gather}
This relation provides a one-to-one map between solutions of equation~(\ref{BogomolnyEq}) and solutions
of the system of equations~(\ref{NahmEq}).  It is a nonlinear version of the Fourier transform called the Nahm transform.

\subsection{The monopole Spectral Curve}

Before we begin, let us recall the notion of the {\em minitwistor space} ${\bf T}$, which, in the case of a f\/lat three-dimensional space, is the space of lines in ${\mathbb R}^3.$  A line in ${\mathbb R}^3$ can be written in the form $\vec{x}=t \vec{n}+\vec{v},$ where $\vec{n}$ is a unit direction vector and $\vec{v}$ is the displacement vector orthogonal to $\vec{n}.$  Each oriented line is uniquely specif\/ied by the pair $(\vec{n}, \vec{v})$ with $|\vec{n}|=1$ and $\vec{v}\cdot\vec{n}=0.$ Thus the space of lines  in ${\mathbb R}^3$ forms  the tangent space to a two-sphere, where $\vec{n}$ specif\/ies the point on the two-sphere and $\vec{v}$ specif\/ies a point in the tangent space at it.  Identifying the two-sphere with the Riemann sphere with a coordinate $\zeta,$ we give the minitwistor space ${\bf T}$ a natural complex structure.  Thus ${\bf T}=T{\mathbb P}^1$ and locally we introduce coordinate $\eta$ in the f\/iber, so that $\eta\frac{\partial}{\partial\zeta}\in T{\mathbb P}^1.$

As proposed by Nigel Hitchin in \cite{Hitchin:1982gh}, to every solution $(A,\Phi)$ of the Bogomolny equation one can associate a scattering problem.  For each line consider a dif\/ferential equation
\begin{equation}\label{scattering}
\left(D_{\vec{n}}+\Phi\right)\Psi=0,
\end{equation}
where $D_{\vec{n}}$ is the covariant derivative along the line and $\Psi$ is a section of the restriction of the bundle $E$ to this line.  For some lines this problem has a bound state, i.e. a square integrable solution $\Psi.$  We call such a line {\em spectral}.   Given a monopole solution of charge $N$ for any unit vector~$\vec{n}$ generically there will be exactly $N$ spectral lines along $\vec{n}.$   Since every line in ${\mathbb R}^3$ corresponds to a point in ${\bf T},$ the set of points corresponding to spectral lines forms a curve ${\bf S}$ in ${\bf T}.$  Thanks to the fact that $(A, \Phi)$ satisf\/ied Bogomolny equation, this curve ${\bf S}$ is holomorphic.  Moreover, since a  point on this curve corresponds to a line in ${\mathbb R}^3$ with a nontrivial square integrable solution $\Psi$ of equation~(\ref{scattering}), we obtain a line bundle ${\bf\cal L}$ over ${\bf S}.$ See \cite{Hitchin:1982gh}  and \cite{Hitchin:Constraint} for details.

So far we have outlined a map which for every monopole $(A,\Phi)$ produces a pair  $({\bf S}, {\bf\cal L})$ of the curve and the bundle.   These satisfy certain conditions formulated below in Section \ref{section:Constraint}.  So long as these conditions are satisf\/ied the map is one-to-one \cite{Hitchin:1982gh, AtiyahHitchin}.  For a monopole of charge $N$ the curve ${\bf S}\subset T{\mathbb P}^1$ is an $N$ fold cover of the base ${\mathbb P}^1.$   In other words, knowing ${\bf S}\subset T{\mathbb P}^1$ and ${\bf\cal L}\rightarrow{\bf S}$ we can, in principle, reconstruct a unique monopole solution (up to a gauge transformation).

\subsection{Nahm Spectral Curve}\label{section:NahmCurve}
The Nahm transform is a one-to-one map, mapping a monopole of charge $N$ to a solution of the Nahm equations where $T_j(s)$ are functions of the variable $s$ with $N\times N$ Hermitian matrix values.  Now we brief\/ly outline how the same curve ${\bf S}$ and the line bundle ${\bf\cal L}$ emerge from the corresponding Nahm data $(T_0(s)$, $T_1(s)$, $T_2(s)$, $T_3(s))$.

Introducing an auxiliary parameter $\zeta$ def\/ine
\begin{gather*}
A=i T_0+ T_3+\zeta(T_1+i T_2),\\
L=-T_1+i T_2+2i\zeta T_3+\zeta^2(T_1+i T_2),
\end{gather*}
then the system of the Nahm equations (\ref{NahmEq}) can be written as
\[
\frac{d}{ds} L+[A,L]=0.
\]
The above equation implies that the eigenvalues of $L$ are independent of $s$,
thus any solution of the Nahm equations def\/ines a spectral curve ${\bf S}\subset {\rm Tot}{\cal O}(2)=T{\mathbb P}^1$ given by
\[
{\bf S}: \det(L-\eta)=0.
\]
Since the curve is the curve of eigenvalues the corresponding eigenspaces form a line bundle~${\bf\cal L}.$ These are exactly the same as the curve and the bundle of the corresponding monopole constructed in the previous section.

\subsection{The Twistor Curve constraints}\label{section:Constraint}
The constraint that every twistor curve satisf\/ies is formulated in terms of a line bundle $L$ over the minitwistor space ${\bf T}.$  In terms of the coordinates $(\zeta, \eta)$ introduced above the line bundle $L^x(m)$ is def\/ined by its transition function $\zeta^{-m}e^{-x\eta/\zeta}$.  There is a natural restriction of this bundle $L|_{S}$ to any curve $S.$
In case of a regular $SU(2)$ monopole of charge $K$ the ``vanishing theorem'' of \cite{Hitchin:Constraint} states that the twistor curve satisf\/ies the following conditions:
\begin{alignat}{3}
&L^x(k-2)\vert_{{\bf S}}\quad  && {\rm is \   nontrivial \  for } \ \  0<x<2, &\nonumber\\
&L^2\vert_{{\bf S}} &&  {\rm is \  trivial}.&\label{Constraint}
\end{alignat}

\section{Two descriptions of the Spectral Curve ${\bf \Sigma}$:\\ the Hitchin system and instantons}

The Hitchin system of equations is written for a Hermitian gauge f\/ield  $A=A_1 dx^1+A_2 dx^2$  and a pair of Higgs f\/ields $\Phi_1$ and $\Phi_2$ on some Riemann surface $X$.  The corresponding connection is $D_j=\partial_j+i A_j$ and the equations are
\begin{gather}
[D_1,D_2]=-[\Phi_1,\Phi_2],\nonumber\\
{}[D_1,\Phi_1]=-[D_2,\Phi_2],\label{HitchinEq}\\
{}[D_1,\Phi_2]=[D_2,\Phi_1] .\nonumber
\end{gather}
The Nahm transform maps any solutions of the Hitchin system  on a torus $T^2$ to a doubly-periodic instanton \cite{Jardim:1999dx, Jardim:2002tf}.

\subsection{The Higgs Curve ${\bf \Sigma}$ from the Hitchin system}\label{section:HiggsCurve}

The Hitchin system can be written in the following form
\begin{gather}
[D,\bar{D}]=-[\Phi,\Phi^{\dagger}],\nonumber\\
{}[\bar{D},\Phi]=0,\label{Eq:HitchinSys}
\end{gather}
if we put $D=D_1-iD_2$, $\Phi=\Phi_1-i\Phi_2,$ so that
$\bar{D}=D_1+iD_2$ and $\Phi^{\dagger}=\Phi_1+i\Phi_2.$
The second one of these, equations (\ref{Eq:HitchinSys}), implies that the curve ${\bf \Sigma}\subset T^*X$
def\/ined by
\[
{\bf \Sigma}: \det(\Phi-w)=0,
\]
is holomorphic. We shall refer to this curve as the {\em Higgs Curve} or the {\em Brane Curve}.
Since ${\bf \Sigma}$ is a~curve of eigenvalues it comes with the line bundle ${\bf\cal N}\rightarrow{\bf \Sigma}.$
Each f\/iber of ${\bf\cal N}$ is the corresponding eigenspace.

Thanks to the Donaldson's theorem \cite{Donaldson} extended to this case $({\bf \Sigma}, {\bf\cal N})$ def\/ine the solution of the Hitchin system uniquely (up to a gauge transformation).

\subsection{The Higgs Curve from doubly-periodic instantons}
\label{section:InstantonCurve}

A doubly-periodic instanton is a self-dual connection on ${\mathbb R}^2\times T^2$.  Let us introduce complex coordinates $z\in{\mathbb C}\simeq{\mathbb R}^2$ and $w\in T^2\simeq{\mathbb C}/({\mathbb Z}\times{\mathbb Z}).$ Instanton equations  can be written as
\begin{gather}
{}[D_z, D_w]=0,\nonumber\\
{}[D_z,\bar{D}_z]=[D_w, \bar{D}_w].\label{InstantonEq}
\end{gather}
As demonstrated in  \cite{Jardim:2002tf}, following the ideas of  \cite{Friedman:1997yq, Friedman:1997ih}, a spectral curve is associated to any solution of these equations in the following manner.   For any point $z\in{\mathbb C}$ we can consider monodromies of a f\/lat connection on the torus $T^2_z$ with the holomorphic combination equal to  $D_w$  of the instanton.  If the instanton bundle is of rank $K,$ then the eigenvalues of these monodromies  correspond to $K$ points in the dual torus $\check{T}^2.$  Thus the set of all these points for all values of $z$ def\/ines a curve ${\bf \Sigma}$ which is a $K$ fold covering of ${\mathbb R}^2$.  Thanks to the fact that the initial connection satisf\/ied the  f\/irst equation in (\ref{InstantonEq}), this curve ${\bf \Sigma}$ is holomorphic.

It is no coincidence that the curves of Section~\ref{section:InstantonCurve} and Section \ref{section:HiggsCurve} are both denoted by ${\bf \Sigma}.$  As proved in \cite{Jardim:2002tf}, if a doubly-periodic instanton (\ref{InstantonEq}) is related via the Nahm transform to a solution of the Hitchin system (\ref{Eq:HitchinSys}) the corresponding curves coincide.

\section{A bridge between ${\bf S}$ and ${\bf \Sigma}$: periodic monopoles}

One might consider various limits in order to relate the two curves.  For example, in a certain limit the Hitchin system degenerates to a system of the Nahm equations and doubly-periodic instanton become a monopole.  In this section, however, we explore the case where the curves of both kinds appear in the same problem.

Let us consider periodic monopoles \cite{Cherkis:2000cj}, i.e. solutions of Bogomolny equation (\ref{BogomolnyEq}) on ${\mathbb R}^2\times S^1.$  What are the corresponding curves in this situation?

\subsection{The Higgs Curve ${\bf \Sigma}$ for periodic monopoles}

Identifying the ${\mathbb R}^2$ of ${\mathbb R}^2\times S^1$ with a complex plane ${\mathbb C}$ with a coordinate $z$ we consider the monodromy $W(z)$ of the modif\/ied connection $D+\Phi$ around the circle $S^1_z.$  In other words, if $w(z,0)=1$ and
\begin{equation}\label{MonodromyEq}
\left(\frac{\partial}{\partial \varphi}+i A_\varphi+\Phi\right) w(z,\phi)=0,
\end{equation}
then $W(z)=w(z, R).$  Here $R$ is the period of the coordinate $\varphi$ along the $S^1$ at $z$.   As argued in~\cite{Cherkis:2000cj}, the curve def\/ined by the following equation is holomorphic.
\[
{\bf \Sigma}_p: \det\left(W(z)-t\right)=0.
\]
Since $t$ is an eigenvalue of a monodromy matrix it is nonzero, thus $t\in{\mathbb C}^*$ and $z\in{\mathbb C},$ and
 ${\bf \Sigma}_p\subset {\mathbb C}\times {\mathbb C}^*.$

\subsection{The Twistor Curve ${\bf S}$ for periodic monopoles}

In order to def\/ine the scattering problem in this case, we start with describing the minitwistor space of ${\mathbb R}^2\times S^1.$  We can view ${\mathbb R}^2\times S^1$ as the quotient of ${\mathbb R}^3$ with respect to integer shifts by a vector $\vec{R}\in{\mathbb R}^3,$ with $R=|\vec{R}|.$  A point in ${\mathbb R}^3$ determines a section of $T{\mathbb P}^1$ in the following way.  The set of all oriented lines passing through $\vec{R}$ forms a sphere.  Each of these lines corresponds to a unique point in ${\bf T}=T{\mathbb P}^1,$ thus $\vec{R}$ corresponds to a section $p_{\vec{R}}$ of the tangent bundle $T{\mathbb P}^1.$  Let $p(\zeta)=-x_1+i x_2+2  x_3 \zeta+(x_1+i x_2)\zeta^2.$ In terms of the local coordinates  $(\zeta, \eta),$ that we def\/ined above, $p_{\vec{R}}: \eta=p(\zeta).$

Shifting ${\mathbb R}^3$ by $\vec{R}$ maps a point $(\zeta_0,\eta_0)$ of $T{\mathbb P}^1$ to $(\zeta_0,\eta_0+p(\zeta_0)).$  Thus the minitwistor space~${\bf T}_p$ of ${\mathbb R}^2\times S^1$ is $T{\mathbb P}^1/{\mathbb Z} p_{\vec{R}}.$   Here the subscript $p$ in ${\bf T}_p$ stands for `periodic'.   Note that the the f\/iber above $\zeta$ such that $p(\zeta)\neq 0$ is ${\mathbb C}/{\mathbb Z}\simeq {\mathbb C}^*,$ while the f\/ibers above  $\zeta_1$ and $\zeta_2$  are ${\mathbb C}_{\zeta_1}={\mathbb C}$ and ${\mathbb C}_{\zeta_2}={\mathbb C}.$    Now we have the minitwistor space ${\bf T}_p$ for ${\mathbb R}^2\times S^1.$  Alas, the space ${\bf T}_p$ is not Hausdorf\/f.  In order to avoid this problem for now, let us consider the compliment of the two f\/ibers above the roots $\zeta_1$ and $\zeta_2$ of $p(\zeta).$  We shall denote it by ${\bf T}_p^{\rm Reg}.$  In other words ${\bf T}_p^{\rm Reg}={\bf T}\setminus ({\mathbb C}_{\zeta_1}\cup {\mathbb C}_{\zeta_2}).$ Without loss of generality we can choose the vector $\vec{R}$ to be along the third axis and work away from $\zeta=0$ and $\zeta=\infty.$

From the point of view of the geodesics on ${\mathbb R}^2\times S^1$ one can see why the roots of $p(\zeta)$ are so special.  A generic geodesic is inf\/initely long and projects to a line in ${\mathbb R}^2.$  There is a set of geodesics, however, that are circles each of which  projects to a point on ${\mathbb R}^2,$ namely, these are the geodesics directed along $\vec{R}.$  Clearly, for any two of the circular geodesics we can f\/ind a long geodesic that is arbitrarily close to both.  Excluding the circular geodesics for now, we avoid this problem.

For any long geodesic we can again consider the scattering problem of Hitchin.  A geodesic is called {\em spectral} if, as in the case of monopoles, it has a square integrable solution $\Psi$ of
\begin{equation}\label{ScatteringEq}
(D_{\vec{n}}+\Phi)\Psi=0.
\end{equation}
 The set of spectral geodesics forms a curve ${\bf S}_p\subset{\bf T}_p^{\rm Reg}.$  This cure is holomorphic, which is a~consequence of the Bogomolny equation.

\section{The Hitchin system}
The Nahm transform of a periodic monopole is a Hitchin system on ${\mathbb R}\times S^1.$  Here we describe the curves ${\bf \Sigma}_p$ and ${\bf S}_p$ in terms of this Hitchin data $(A,\Phi)$.

The Higgs curve ${\bf \Sigma}_p$ of the eigenvalues of $\Phi_1-i\Phi_2$  is def\/ined in the same way as for the general Hitchin system of Section \ref{section:HiggsCurve}.  In our case the Hitchin data is def\/ined on ${\mathbb R}\times S^1\simeq {\mathbb C}^*$ and thus the resulting curve lies in $T^*{\mathbb C}^*={\mathbb C}\times{\mathbb C}^*.$

In order to def\/ine the Twistor Curve ${\bf S}_p$  let us follow  \cite{LambdaConnection}
 and introduce the so-called $\lambda$-con\-nection
\[
D_\zeta=D+\zeta\Phi,\qquad  \bar{D}_\zeta=\bar{D}+\frac{1}{\zeta}\Phi^{\dagger},
\]
where the auxiliary parameter $\zeta\in{\mathbb C}^*.$ Then the Hitchin equations (\ref{HitchinEq}) are equivalent  to f\/latness of this connection: $[D_\zeta, \bar{D}_\zeta]=0$ for all values of $\zeta\in{\mathbb C}^*.$  The monodromy of a f\/lat connection on ${\mathbb R}\times S^1$ along a closed path winding once around the $S^1$ has eigenvalues independent of the choice of the path.  Thus, for any nonzero $\zeta$ we obtain a set of points in ${\mathbb C}^*.$ This def\/ines the curve ${\bf S}_p\subset{\bf T}_p^{\rm Reg}.$

 Clearly the curve ${\bf S}_p$ and the bundle ${\bf\cal L}$ thus obtained satisfy some nontrivial conditions.  In particular, in the limit of the zero radius of the $S^1$ factor of the ${\mathbb R}\times S^1,$ the Hitchin system degenerates into the system of the Nahm equations.  In this limit ${\bf S}_p$ becomes ${\bf S}$ of Section \ref{section:NahmCurve}.  Thus whatever the conditions the curve ${\bf S}_p$ has to satisfy should degenerate to the constraint of equation~(\ref{Constraint}).

\section{The question}
For a generic periodic monopole, we expect the same theorems to hold as in the case of a~monopole in ${\mathbb R}^3$ and in the case of doubly-periodic instantons.  Namely, each pair $({\bf \Sigma}_p, {\bf\cal N})$ of the curve ${\bf \Sigma}_p$ and a line bundle ${\bf\cal N}$ is in one-to-one correspondence with a periodic monopole.  On the other hand, each pair $({\bf S}_p, {\bf\cal L})$ satisfying certain conditions is also in one-to-one correspondence with a periodic monopole.  Thus there should be a one-to-one map mapping a pair $({\bf \Sigma}_p, {\bf\cal N})$ to the pair $({\bf S}_p, {\bf\cal L}).$

Since the Nahm transform relates periodic monopoles with the solutions of the Hitchin system on ${\mathbb R}\times S^1$ \cite{Cherkis:2000ft, Cherkis:2000cj}, periodic monopoles in the above argument can be substituted with that Hitchin system.

Now we are in a position to formulate the question: what is the explicit map relating the Higgs curve ${\bf \Sigma}$ to the Twistor Curve ${\bf S}$?

\pdfbookmark[1]{Appendix. Speculations}{appendix}
\section*{Appendix. Speculations}

Let us come back to the minitwistor space ${\bf T}_p$.  The curve ${\bf S}_p$ was def\/ined on the compliment of the `problematic' f\/ibers ${\mathbb C}_{\zeta=0}$ and ${\mathbb C}_{\zeta=\infty}.$  The f\/iber ${\mathbb C}_{\zeta=0}$ can be identif\/ied with the ${\mathbb R}^2$ factor of the space ${\mathbb R}^2\times S^1$ on which the periodic monopole is def\/ined.  (The f\/iber ${\mathbb C}_{\zeta=\infty}$ is identif\/ied with the same ${\mathbb R}^2$ factor but with the opposite orientation.)  Each point in this f\/iber corresponds to a closed geodesic $S^1$.  It is exactly these geodesics that are used to def\/ine the Higgs Curve~${\bf \Sigma}_p.$  Moreover, one cannot fail to notice that the operator in equation~(\ref{MonodromyEq}) is exactly the same as in the scattering problem equation~(\ref{ScatteringEq}).  (Note, however, that $\Psi$ in equation~(\ref{ScatteringEq}) is a section, while $w$ in equation~(\ref{MonodromyEq}) is a parallel transport operator.)   Thus, in some sense, the curve ${\bf \Sigma}_p$ (or rather its zeros with $t=0$) is a limit of ${\bf S}_p$ at the special f\/iber at $\zeta=0$ (and $\zeta=\infty)$.

These is a potentially important dif\/ference between the role the curves ${\bf \Sigma}$ and ${\bf S}_p$ play in the integrable system. A particular doubly-periodic monopole solution (or a corresponding solution to the Hitchin system) is determined in terms of a curve ${\bf \Sigma}$ and a point in its Jacobian ${\rm Jac}({\bf \Sigma})$, while in terms of the curve ${\bf S}_p$ such solution is determined by ${\bf S}_p$ and a linear f\/low on its Jacobian ${\rm Jac}({\bf S}_p).$  These two cases correspond to a two dif\/ferent views on the monopole (or Hitchin equations). In the earlier case the solution is viewed as a point in the conf\/iguration space of an integrable system. In the latter case, the same solution is interpreted as  an evolution trajectory of an integrable system.

\subsection*{Acknowledgments}   We thank  Pierre Deligne, Tamas Hausel, Nigel Hitchin, Anton Kapustin, Lionel Mason, Tony Pantev, Emma Previato, Samson Shatashvili, and Edward Witten for useful discussions.

This work is supported by the Science Foundation Ireland Grant No. 06/RFP/MAT050 and by the European Commision FP6 program MRTN-CT-2004-005104.

\pdfbookmark[1]{References}{ref}
\LastPageEnding
\end{document}